\documentstyle[12pt]{article}

\newcommand{\be}{\begin{equation}}
\newcommand{\ee}{\end{equation}}
\newcommand{\bea}{\begin{eqnarray}}
\newcommand{\eea}{\end{eqnarray}}

\newcommand{\p}[1]{(\ref{#1})}
\newcommand{\ba}{\begin{array}}
\newcommand{\ea}{\end{array}}
\newcommand{\vs}[1]{\vspace{#1 mm}}

\def\bbox{{\,\lower0.9pt\vbox{\hrule \hbox{\vrule height 0.2 cm
\hskip 0.2 cm \vrule height 0.2 cm}\hrule}\,}}
\newcommand{\dsl}{\pa \kern-0.5em /}

\newcommand{\pa}{\partial}

\begin{document}

\topmargin 0pt
\oddsidemargin 5mm

\renewcommand{\thefootnote}{\fnsymbol{footnote}}
\begin{titlepage}

\setcounter{page}{0}

\rightline{\small \hfill hep-th/0003197}
\rightline{\small \hfill LPT-ENS 00/15}
\rightline{\small \hfill DAMTP 2000-66}
\vskip -.75em\rightline{\small \hfill }
\vskip -.75em\rightline{\small \hfill June 2000}

\vs{10}
\begin{center}
{\Large Domain Walls and Solitons in Odd Dimensions}\\
\vs{10}
G.W. Gibbons\footnote{Permanent address: DAMTP, Cambridge University}\\
Yukawa Institute for Theoretical Physics,\\
University of Kyoto,\\
Sakyo-ku,\\
Kyoto 606-8502, Japan\\
\vs{2}
and\\
\vs{2}
N.D. Lambert\footnote{Current address: Dept. of Mathematics, King's College
London}\\
Laboratoire de Physique Th{\'e}orique CNRS-ENS\\ 
24 rue Lhomond, F-75231 Paris Cedex 05,\\ 
France\footnote{UMR 8549, Unit{\'e}   Mixte de Recherche du 
Centre National de la Recherche Scientifique et de 
l' Ecole Normale Sup{\'e}rieure.}
\vs{5}
{\em }
\end{center}
\vs{7}
\centerline{{\bf Abstract}}
We discuss the
existence of smooth soliton solutions which interpolate 
between supersymmetric vacua in odd-dimensional theories. 
In particular we 
apply this analysis to a wide class of  supergravities to 
argue against the existence of smooth  domain walls
interpolating between supersymmetric vacua. We find that if the
superpotential changes sign then any Goldstino modes will diverge.

\end{titlepage}
\newpage
\renewcommand{\thefootnote}{\arabic{footnote}}
\setcounter{footnote}{0}

\section{Introduction}

Soliton solutions are a central theme in the study of supersymmetric
theories.  In particular domain wall solitons in five-dimensions have
received  substantial attention because of their potential to admit chiral
fermion zero modes.
In this paper we wish to describe some 
observations about chiral fermion zero modes of domain walls viewed as stable 
finite energy solutions which interpolate
between two vacua of the theory. Indeed we will argue, at least in a 
wide class of supersymmetric  theories in odd dimensions, 
that no smooth domain walls exist.

The  analysis given
below  was initiated by the question as to  whether
or not a Randall-Sundrum scenario \cite{RS} can be extended to a 
smooth domain
wall in a supergravity theory. This question has several
motivations. 
It was pointed out in \cite{BC} that
such an embedding would solve the fine-tuning problem associated with
matching the domain wall tension and bulk cosmological constant needed in
\cite{RS}. Indeed
without supersymmetry one is led to question the general stability of a domain
wall \cite{CG}. In addition, with a smooth domain wall solution 
one can improve upon the thin wall approximation in
\cite{RS}
and provide a complete non-linear analysis of the Randall-Sundrum
scenario \cite{CG}. Finally there is widespread belief that supersymmetry and
supergravity are relevant phenomenologically and in this context it is
natural to embed our universe in a  higher dimensional theory 
containing supergravity.
Certainly from a theoretical point of view one would like to place
such a ``brane-world'' in the context of supergravity and ultimately 
string theory. 
The difficulty in obtaining a smooth Randall-Sundrum domain wall in
five-dimensional supergravity has been discussed recently
\cite{CLP,KL,BC2} and a no-go theorem can be proven in various  
cases \cite{KL,BC2,KLS,G}\footnote{Recently the original but discontinuous
Randall-Sundrum domain wall has been embedded into a 
supergravity \cite{ABN}.}. 
In this paper we argue against the existence of smooth
domain walls interpolating between supersymmetric vacua  
on rather general grounds in a wide class of
odd-dimensional supergravities (although not all, e.g. see \cite{FGPW}).

Supersymmetric domain wall spacetimes have also received interest
recently due to their role in the AdS/CFT correspondence. In
particular the domain wall central charge has been identified with the 
c-function  of a four-dimensional field theory \cite{FGPW,GPPZ}. From
this perspective the absence of smooth Randall-Sundrum
domain walls in a
particular supergravity is 
interpreted as the statement that the (monotonic) c-function \cite{GPPZ,FGPW}
\be
C(r) = {C_0\over |W(r)|^{D-2}}\ ,
\label{cfunction}
\ee 
is bounded along the  renormalisation group flow (i.e. that 
$W(r)$ does not pass through zero).

The obstruction to finding supergravity domain walls in five
dimensions seems to be obtaining solutions where the real
superpotential $W$ changes sign \cite{CLP,KL,BC2,KLS}. 
In supergravity theories $W$ appears
in the mass terms for the fermions. Domain walls that interpolate between
regions in which  $W$ changes sign
connect regions with positive fermion mass to those with negative
fermion mass.
In even dimensions the sign of a fermion mass term has no physical 
significance; it may be reversed by multiplying the fermion field by
$\Gamma^{D+1}$. By contrast, in odd dimensions, $\Gamma^{D+1}=\pm 1$,
and a fermion mass term breaks parity. Furthermore in odd dimensions there are
two inequivalent irreducible representations of the Clifford algebra
labelled by the sign of $\Gamma^{D+1}$. Theories with different signs
for
the $\Gamma$-matrices are rather like different superselection sectors
and one would not expect that these two sectors could be
realised in single connected spacetime. However a
change in sign of all
fermion masses may be effected by a change in sign of the $\Gamma$-matrices.
Thus a domain wall in an 
odd-dimensional theory in which $W$ changes sign looks as if it
connects
two distinct superselection sectors and one might doubt that this is
physically sensible.
Perhaps this is the reason why domain walls with
$W$ changing sign have not been found in supergravity.  A similar
reservation was raised in \cite{KL}. 

This also raises the
question of whether domain walls coupled to fermions in which mass
terms
change sign can exist in flat space theories, supersymmetric or not. 
As is well-known such domain
walls admit localised zero energy fermion modes which are
chiral. As long as the worldvolume theory is not anomalous, this would
seem to lead to no contradiction. Of course if it were anomalous,
there
would have to be some in-flow to balance the anomaly and such domain
walls might be incompatible with being supersymmetric, i.e. BPS.

Thus it seems that the key
to  understanding the absence of supersymmetric domain wall solutions lies
in understanding the Goldstino fermions.
Therefore the rest of this paper is organised as follows. In 
section two we shall discuss non-gravitational domain
walls. We find that even if four-dimensional 
supersymmetric domain walls exist 
their Goldstinos are non-chiral and hence non-anomalous. 
Finally in section three we discuss domain walls in
supergravity. There we find that 
the Goldstino fermion modes would diverge if $W$ changed sign.

\section{Non-Gravitational Domain Walls}

\subsection{Fermion Zero-Modes}

Let us consider a general theory which includes a fermionic field
$\psi$.  Around
any vacuum of this theory we may consider the fluctuations
of the fermion which we assume satisfy the Dirac equation (we 
use a ``mostly'' plus metric in $D$ spacetime
dimensions, $m,n=0,1,2,...,D-1$)
\be
\Gamma^{m}\nabla_m \psi + M\psi =0 \ .
\label{chieq}
\ee
As is well known this equation admits both positive and negative energy
solutions $\psi^{(\pm)}$. 
In particular, particles at rest have one-particle wave functions
given by ``plane-wave'' solutions
$\psi^{(\pm)} = e^{\mp i|M|t}\eta_{\pm}$ where $\eta_{\pm}$ 
is a constant spinor and
$i\Gamma^0 \eta_{\pm} = \pm sign(M) \eta_{\pm}$.
The resolution of this ``energy crisis'' in the quantum
theory is to simply assert that in a given vacuum all the negative
energy states $\psi^{(-)}$ are filled. 

Now imagine that there is a domain solution associated with a
scalar field $\phi(r)$ which interpolates between
two vacua, where $r$ is the coordinate transverse to the domain wall. 
As a consequence the fermion mass $M(r)$ becomes dependent upon
$r$. We must now look for solutions of the form
\be
\psi = e^{ip_\mu x^\mu}\chi(r)\ ,
\ee
where $\mu=0,1,2,...,D-2$. There are two cases to consider. In the
first case $\Gamma^\mu p_\mu\chi=0$ so that $p_\mu p^\mu=0$. 
The solution then takes the form
\be
\chi(r) = 
\chi_{\pm}e^{ip_\mu x^\mu} exp\left( \mp \int_0^r M(r')dr' \right)\ ,
\label{masslessmode}
\ee
where $\Gamma^r\chi_\pm=\pm \chi_\pm$. If the spacetime dimension is
{\sl odd}, $\Gamma^0\Gamma^1\ldots \Gamma^r=\pm 1$ and hence $\Gamma^r$
determines the chirality of the fermions with respect to the wall.
If $M(r)$ changes sign then one
chiral mode in \p{masslessmode} will be normalisable and the other 
non-normalisable.
This implies that a massless chiral fermion is bound to the domain wall. 
If $M(r)$ does not change sign then neither mode is normalisable.
We will not be interested in this situation in this paper. 

In the case that $\Gamma^\mu p_\mu \chi\ne 0$ we may, without loss of
generality, take $p_\mu=(-E,0,\ldots ,0)$. Using the
$\Gamma$-matrix algebra we have $i\Gamma^0\chi_\pm = \chi_\mp$ leading to
the following system of equations
\bea
(\partial_r + M)\chi_+ &=& E\chi_-\ ,\nonumber\\
(\partial_r - M)\chi_- &=& -E\chi_+\ .\nonumber\\
\label{coupledeq}
\eea
Elimination leads to a second order differential equation in which $\chi_\pm$
decouple
\be
\left(-\partial_r^2+V_\pm (r)\right)\chi_\pm = E^2\chi_\pm\ ,\quad
V_\pm (r) = M^2(r) \mp \partial_r M(r)\ .
\ee
For an explicit 
example we let $M(r) = m {\rm tanh}(mr)$. We can solve exactly
these equations since $V_+ = m^2$ and hence
\be
\chi_+ = Ae^{i\sqrt{E^2-m^2}\ r} +B e^{-i\sqrt{E^2-m^2}\ r}\ ,
\ee
where $A$ and $B$ are arbitrary constants. These represent plane wave
solutions for $E^2 \ge m^2$ but are non-normalisable if $E^2 < m^2$.
We may then obtain 
\be
\chi_- = E^{-1}(\partial_r + m{\rm tanh}(mr))\chi_+\ .
\label{flip}
\ee
This completes the spectrum of fermion modes in the domain wall 
background. In summary, if $M(r)$ changes sign then there is a single chiral
fermion with zero energy localised on the wall. 
In addition there are modes with non-vanishing energy. 
Note that we may perform a boost to transform these modes into their 
rest frame where $E=m$ so that  the fermion wave function is
\be
\psi = e^{-imt} (\chi^0_+ + {\rm tanh}(mr)\chi^0_- )\ ,
\ee
which smoothly interpolates between the constant eigenstates
$\eta^0_{\pm} = \chi^0_+ \pm \chi^0_-$ of $i\Gamma^0$ that appear in the
Dirac sea in each vacuum.

This example shows that one can have domain walls in odd dimensions 
which interpolate between two vacua with opposite signs for the
fermion
mass. Thus if the domain wall has a ${\bf Z}_2$ symmetry then globally
parity is a good symmetry, even though it is broken in each vacua.
We also see that fermions with positive energy can travel freely between
the two vacua. In particular there is no mixing of positive and
negative frequencies which would indicate a quantum instability.

\subsection{Chiral Goldstinos}

Let us now consider supersymmetric domain walls in five dimensions
where the fermion zero modes arise as Goldstinos.
First we note that
the mass term in \p{chieq} comes from a term of the form 
$M(\phi)\bar\psi\psi$ in the Lagrangian. For a supersymmetric theory in 
five dimensions, the scalar
$\phi$ belongs in a vector multiplet, tensor multiplet or a hyper multiplet.
For a vector multiplet a coupling of the form $\phi\bar\psi\psi$ might seem
natural, since this comes from the dimensional reduction of a covariant
derivative term for $\psi$ in six dimensions. However in this case there
can be no potential for $\phi$ by gauge invariance and hence no domain wall.
In fact in general, if a five-dimensional multiplet with eight supercharges 
comes via compactification from six dimensions, 
then it must come from a chiral 
multiplet. Therefore there are no fermion mass terms
possible  (unless we consider theories with sixteen 
supercharges in which case there are no scalar potentials possible).

In fact regardless of which kind of multiplet $\phi$ belongs in 
five dimensions,
when solving
the  domain wall equations we
effectively compactify the system to the  two dimensions $t$ and $r$. The
action will then have $(4,4)$ supersymmetry and this constrains the potential
to take the form $V = g_{ij}k^ik^j$ where $g_{ij}(\phi)$ 
is a hyper-K\"ahler metric 
appearing in the scalar kinetic term and $k^i(\phi)$ is a tri-holomorphic
Killing vector. The Yukawa term is 
$\nabla_i k_j(\phi)\bar\psi^i\Gamma^{0r}\psi^j$ \cite{AF}.
Let us again write  $\Gamma^r\psi^i_\pm = \pm \psi^i_\pm$,   
$i\Gamma^0\psi^i_\pm = \psi^i_\mp$ and $\psi^i = e^{-iEt}\chi^i$. 
In this case the equations of motion for the fermions become
\bea
\nabla_r \chi^i_+ - M^i_{\ j}\chi^j_- &=& E\chi^i_- \ ,\nonumber\\
\nabla_r \chi^i_- - M^i_{\ j}\chi^j_+ &=& -E\chi^i_+ \ ,\nonumber\\
\label{eqtwo}
\eea
where $M^i_{\ j} = \nabla^ik_j$,   
$\nabla_r\psi^i_\pm = \partial_r\psi^i_\pm 
+ \Gamma^i_{jk}\partial_r\phi^j\psi^k_\pm$ and we have ignored a term  cubic 
in the fermions involving the curvature of the metric $g_{ij}$.

The  system \p{eqtwo} is quite different to \p{coupledeq} because the
left hand side contains both $\chi^i_-$ and $\chi_+^i$. 
In particular if $E=0$,  
$\chi^i_+$ and $\chi^i_-$ do not decouple. Moreover if $E=0$, 
$\chi^i_+$ and $\chi^i_-$
satisfy the same equation and hence the fermion
zero modes come in pairs containing both chiralities. 
Note that this result no longer holds in lower dimensions since  the
resulting two-dimensional transverse theories need only have $(1,1)$ or $(2,2)$
supersymmetry and equations of motion of the form  $\p{coupledeq}$
are possible \cite{AF}. We also observe
that this form of the Yukawa term does not break parity.

\section{Domain Walls in Supergravity}

We now wish to discuss  domain wall solitons in 
supergravity.  First let us review some basic features of supergravity
domain walls. We assume that the bosonic  action takes the form
\be
S = \int d^D x \sqrt{-g} \left( R 
- \gamma_{AB}(\phi)\partial_m \phi^A\partial^m\phi^B
- V(\phi)\right)\ ,
\label{action}
\ee
where $\phi^A$, $A = 1,2,3,...,N$ are scalar modes and we assume that
metric $\gamma_{AB}$ appearing in their kinetic term is positive definite. 
We further assume that \p{action} is the consistent truncation of a
supergravity theory which is invariant under supersymmetry 
transformations of the form
\bea
\delta \psi_m &=& (\nabla_m \epsilon + W\Gamma_m\epsilon 
+ \partial_mW_2\epsilon)\ , \nonumber\\ 
\delta\lambda_A &=& (-{1\over2}\gamma_{AB}\Gamma^m\partial_m\phi^B +
W_{3A})\epsilon\ .\nonumber\\ 
\label{susy}
\eea
Here $W,W_2$ and $W_{3A}$ are functions of the scalars
$\phi^A$
which we will avoid specifying in order to keep our argument as general
as possible.  In fact we can 
remove the term in \p{susy} involving $W_2$ by performing the field
redefinitions $\epsilon \rightarrow e^{-W_2}\epsilon$, 
$\psi_m \rightarrow e^{-W_2}\psi_m$ and
$\lambda_A \rightarrow e^{-W_2}\lambda_A$. Therefore,
without loss of generality, we set $W_2=0$.
We have also
assumed that any internal indices on the spinors $\epsilon$ may be
ignored. This form for the supersymmetry transformation 
is quite general for $N=2$
supergravity in five dimensions but does not include all 
extended supergravities (e.g. see \cite{FGPW}). We will also ignore
any higher order fermion terms since it is clear that their inclusion
would not affect our discussion.

Let us now look for a supersymmetric domain wall. 
Without loss of generality we may choose the spacetime to
have the metric
\be
ds^2 = dr^2 + e^{2A(r)}\eta_{\mu\nu}dx^\mu dx^\nu\ ,
\label{metric}
\ee
where $\mu,\nu=0,1,2,...,D-2$ and the scalars depend only on
$r$. The requirement that some supersymmetry is preserved gives rise
to the Bogomoln'yi equations
\bea
A' = \mp 2 W\ ,\nonumber\\
{\phi^A}' = \pm 2\gamma^{AB} W_{3B}\ ,\nonumber\\
\label{bogomolnyi}
\eea
where a prime denotes differentiation with respect to $r$. The
preserved supersymmetries (i.e. Killing Spinors) for these domain walls are
\be
\epsilon = e^{{1\over2}A}\epsilon_\pm\ ,
\ee
where $\Gamma_{\underline r}\epsilon_\pm=\pm\epsilon_\pm$ 
and an underlined index refers to the tangent frame. 

It is instructive to  consider supersymmetric vacua of this
theory. Here we set all the scalars to constants $\phi^A=\phi^A_0$. 
Clearly this can only occur
at the ``critical'' points where $W_{3A}(\phi^A_0)=\partial 
V/\partial\phi^A=0$. The 
spacetime \p{metric} is now just pure AdS space with 
$A = \mp2W(\phi^A_0)r$. In this case there are additional
Killing Spinors given by
\be
\epsilon = \left(e^{-{1\over2}A} -
2W(\phi_0^A)e^{{1\over2}A}x^\nu\Gamma_{\underline \nu}\
\right)\epsilon_\mp\ .
\ee
There may also be non-supersymmetric vacua where $\partial
V/\partial\phi^A=0$ but $W_{3A}\ne 0$. However we will have little so
say about these cases.

In a Randall-Sundrum domain wall $A(r) \sim -|r|$ as $r\rightarrow
\pm\infty$ \cite{RS}.
Thus asymptotically $g_{00} = e^{2A}$ falls off exponentially and gravity is
is localised to the domain wall. This will be the case for a
domain wall of the theory \p{action} if $W$ changes sign between the
two vacua. For example in the original proposal \cite{RS} 
there are no scalars $\phi^A$ or fermions $\lambda_A$ and 
$V\sim -W^2$
is constant. The domain wall is obtained by simply choosing the sign of
$W$ to be positive on one side and negative on the other,
i.e. $W(r)$ is discontinuous. Note that from the point of view of
the supergravity equations this domain wall is
equivalent to keeping $W$ fixed everywhere but choosing one
representation for the $\Gamma$-matrices on one side and the opposite
representation (obtained by $\Gamma^m \rightarrow -\Gamma^m$) 
on the other. Given the comments in
the introduction it is natural to be concerned  that this is unphysical.

In supergravity theories there is a standard argument for the
stability of BPS backgrounds. We will briefly review it here and note
that it is insensitive to a change in sign of $W$. 
We construct a ``Nester'' tensor \cite{N}
\be
N^{mn} = \bar\epsilon \Gamma^{mnp}\delta\psi_p ,
\label{Nester}
\ee
with $\delta \psi_m$ given in \p{susy}. Such a tensor has the
property that, on shell, 
\be
\nabla_m N^{mn} =  \bar{{\delta\psi_m}}\Gamma^{mnp}\delta \psi_p
+ \gamma^{AB} \bar{{\delta\lambda_A}}\Gamma^n\delta\lambda_B\ .
\label{Ncondition}
\ee
So in particular $\nabla_mN^{m0}$ is negative definite (provided that
we impose the Witten condition $\Gamma^m\delta \psi_m=0$ \cite{W}) 
and vanishes if and
only if some supersymmetry is preserved. 
In our case this case the
requirement 
that $N^{mn}$ satisfies \p{Ncondition} implies \cite{T}
\bea
W_{3A} &=& (D-2){\partial W\over \partial \phi^A}\ ,\nonumber\\
V &=& 4(D-2)^2\left[
\gamma^{AB}{\partial W\over \partial\phi^A}{\partial W\over
\partial\phi^B}
-\left({D-1\over D-2}\right)W^2
\right]\ .\nonumber\\
\label{restrictions}
\eea

The Nester
tensor can be used to provide a bound on the tension of an arbitrary domain
wall in terms of a central charge of the supersymmetry algebra which
in turn provides a non-perturbative proof of the stability of the
solution.  Following \cite{CGR,BC} we 
integrate $N^{mn}$ over a spacelike boundary
which encloses the domain wall
\be
{1\over2}\int d\Sigma_{mn} N^{mn} = \int d\Sigma_{0r} N^{0r}
= -\int d\Sigma_0 \nabla_m N^{m0} 
\ge 0 \ .
\label{energybound}
\ee
On the other
hand we can directly evaluate the surface integral 
\be
\int d\Sigma_{0r} N^{0r} = \sigma - |W(r=\infty) - W(r=-\infty)|\ ,
\label{tensionbound}
\ee
where $\sigma$ is the tension of the domain wall and we have assumed that
the domain wall interpolates smoothly 
between two AdS vacua. Combining these two
equations we learn that $\sigma \ge |W(r=\infty) - W(r=-\infty)|$
for all domain walls with equality if and only if some supersymmetry
is preserved.

Note that this proof does not actually require that the
action \p{action} admit  a supersymmetric completion. The proof of
stability merely requires that the identities 
\p{restrictions} hold and that there are solutions to the
supersymmetry Killing spinor equations \p{susy}. In particular it
places
no restriction on the function $W$ and hence any 
domain wall satisfying  \p{bogomolnyi} will be stable in the 
purely bosonic theory \cite{B}.
On the other hand we will shortly
see that some choices of the function $W(\phi)$ can never appear in
a consistent supergravity because one could not consistently
couple the theory to fermions.

To begin our discussion of the fermions   we first obtain their 
equations of motion  by constructing the most general
form  and then imposing
the condition that their variation under
supersymmetry vanishes when the scalars are on-shell. After a lengthy
calculation we find
\bea
\Gamma^m\nabla_m\lambda_A &+& M_{A}^{\ B}\lambda_B
- (D-2)\gamma^{CD}{\partial\gamma_{BD}\over\partial\phi^A} 
{\partial W\over\partial\phi^C}\lambda^B
-{1\over2}{\partial\gamma_{BD}\over\partial\phi^A}
\Gamma^m\nabla_m\phi^B\lambda^D\nonumber\\
&+& {1\over2}\gamma_{AB}\Gamma^m\Gamma^n\nabla_n\phi^B\psi_m - 
(D-2){\partial W\over\partial\phi^A}\Gamma^m\psi_m
=0\ ,
\label{lambdaeq}
\eea
\bea
\Gamma^{mnp}\nabla_n\psi_p &-& (D-2)W\Gamma^{mn}\psi_n +
(D-2){\partial W\over \partial \phi^A}\Gamma^m\lambda^A\nonumber\\
&+&{1\over2}(g^{mn}-\Gamma^{mn})\nabla_n\phi^A\lambda_A=0\ ,\nonumber\\
\label{psieq}
\eea
where
\be
M_A^{\ B} = 2(D-2) {\partial W\over \partial \phi^A\partial \phi^C}
\gamma^{BC} - (D-2)W\delta_A^{\ B}\ .
\ee
Therefore, in a supersymmetric AdS vacuum, we may set $\psi_m=0$ and obtain
the equation of motion
\be
\Gamma^m\nabla_m \lambda_A + M_{A}^{\ B}\lambda_B =0\ .
\label{Mass}
\ee

Consider now a stable domain wall, i.e. one that satisfies \p{bogomolnyi}.
In particular since half of the supersymmetries are
broken, one expects that a finite tension domain wall 
has massless Goldstino $\lambda_A$ modes bound to it.  
Therefore we should look for
a solutions to the fermion equations in a background given by \p{bogomolnyi}
which are invariant under
the Poincare symmetry of the domain wall, i.e. with
$\partial_\mu=\psi_\mu=0$. 
It is important to note that the two fermion equations
\p{lambdaeq} and \p{psieq} 
do not
decouple in this case and we must have $\psi_r\ne 0$. 
Specifically 
we find from  the $m=r$ component of the $\psi_m$ equation \p{psieq}
that $\Gamma^r \lambda_A=\mp\lambda_A$. 
For $m\ne r$ the $\psi_m$ equation implies that
\be
\psi_r = \mp {1\over W}{\partial W\over \partial \phi^A}\lambda^A\ .
\label{psilambda}
\ee
Thus the fermion zero modes are chiral, as expected from the 
chiral form of  the  broken supersymmetries.
Substituting \p{psilambda} into the $\lambda_A$ equation \p{lambdaeq}
and using  
\p{bogomolnyi} yields the equation
\be
\pm\partial_r \lambda_A = W\lambda_A 
+2(D-2)\left(
{\partial^2 W\over \partial\phi^A\partial\phi^B}
 - {1\over W}{\partial W\over \partial\phi^A}
{\partial W\over \partial\phi^B}
\right)\lambda^B\ .
\ee
Thus we obtain the wavefunctions for the chiral Goldstino modes
\bea
\lambda_A &=& {1\over W} {\partial W\over \partial\phi^A}
e^{-{1\over2}A}\epsilon_{\mp}\ ,
\nonumber\\
\psi_r &=& \mp {1\over W^2} \gamma^{AB}
{\partial W\over \partial\phi^A}{\partial W\over \partial\phi^B}
e^{-{1\over2}A}\epsilon_{\mp}\ ,\nonumber\\
\label{Goldstinos}
\eea
where $\epsilon_{\mp}$ is
a constant spinor satisfying $\Gamma^r\epsilon_{\mp}=\mp\epsilon_{\mp}$.
From \p{Goldstinos} it is
clear that if $W(r)$ passes through zero (e.g. if $W$ changes sign)
the Goldstino modes will diverge on the domain wall (or more precisely where
$W=0$) and will not be normalisable. From supersymmetry we expect that
a smooth finite tension domain wall should have Goldstinos. We therefore
conclude that $W$ can not change sign (by passing through zero)
in a supergravity theory.  In particular there are no smooth domain walls
of the Randall-Sundrum type. 

To be more explicit consider a single scalar and suppose that near the point 
where $W$ changes sign we
may write $W \sim (\phi-\phi_0)^\gamma$ with $\gamma \le 1$ so that
$\phi_0$ is not a critical point.
We then find that 
\be
\phi-\phi_0 \sim (r-r_0)^{{1\over 2-\gamma}}\ ,\quad 
A \sim (r-r_0)^{{2\over 2-\gamma}}\ , \quad
\lambda \sim (r-r_0)^{-{1\over 2-\gamma}}\ .
\ee
Thus the Goldstinos diverge where $W$ changes sign.
Note that the metric and Killing spinors are 
bounded near $r=r_0$ but they will
have a cusp singularity for $\gamma<0$ (i.e. if $W$ diverges).
In addition the norm 
$\int dr \sqrt{-g}\bar\lambda\lambda$ will be convergent at $r=0$ 
only for $\gamma<0$.

The argument just given depends crucially on the form of the
supersymmetry transformation rules \p{susy}. In four dimensions, for
example, other possibilities arise and our results on the divergence of
the Goldstino modes will not necessarily apply. Indeed
four-dimensional supersymmetric
supergravity domain walls do exist  \cite{CGR}.

To illustrate the above points we may consider a case with just one scalar 
and a superpotential of the form
\be
W(\phi) = \alpha\left(\phi - {1\over3}\beta^2\phi^3\right)\ ,
\ee
where $\alpha$ and $\beta$ are constants and $\gamma_{AB}=\delta_{AB}$.
The critical points occur
at $\phi_0=\pm \beta^{-1}$ where $W = \pm 2\alpha/3$ 
and indeed one can find smooth
supersymmetric domain walls \cite{DFGK,Gremm,KL}. The stability 
of these domain walls in the bosonic
theory follows from the equations \p{energybound} and \p{tensionbound}. 
However we see  that  
this superpotential can never be consistently embedded
in a supergravity because $W$ changes sign between the two
critical points.

\section{Conclusion}

In this paper 
we have discussed the existence of domain walls in supersymmetric
odd-dimensional theories. In the case of global supersymmetry we
argued that no supersymmetric domain walls exist with purely chiral
Goldstino modes. In the case of supergravities in odd dimensions 
we argued that the superpotential $W$ cannot change sign because if
it did the Goldstino modes would diverge.

\section{Acknowledgments}

This work was initiated during
a visit by G.W.G. to l' Ecole Normale Sup\'erieure.
He would like to thank members of the L.P.T. for
their kind hospitality. He would also like to thank Renata Kallosh,  
Andrei Linde and Shoichi Ichinose for helpful discussions.
N.D.L. was supported by the EU grant ERBFMRX-CT96-0012 and would like
to thank Eug\`ene Cremmer for discussions.  We would
also like to Juan Maldacena for helpful comments on 
original version. 

%%%%%%%%%%%%%%%%%%%%%%%%%%%%%%%%%%%%%%%%%%%%%%%%%%%%%%%%%%%%%%%%%%%%%%%%%%%%%

\end{document}